# Production of $^{177}$Lu at the IFMIF-DONES Facility

Francisco Garcia-Infantes[1], Javier Praena[1], Laura Fernandez-Maza[2], Fernando Arias de Saavedra[1], Ignacio Porras[1]

[1]Universidad de Granada, Granada, Spain    [2]Hospital Virgen de la Arrixaca, Murcia, Spain.

## INTRODUCTION

Since 2016, the project of the International Fusion Materials Irradiation Facility - Demo Oriented NEutron Source (IFMIF-DONES) has been accelerated in its final implementation. In the last years, the most important news have been: the selection of Granada (Spain) as European city host, associated with a Preparatory Phase project [1], and the election of the facility as a key infrastructure in energy by ESFRI (European Strategy Forum on Research Infrastructures) [2]. IFMIF-DONES is the consequence of the urgency in the need of data for the design of DEMO (DEMOstration Power Plant), the future first fusion reactor providing electricity. IFMIF-DONES will be dedicated to the irradiation of materials and alloys in similar conditions of neutron fluence and neutron energy to DEMO [3]. The accelerator planned for IFMIF-DONES will accelerate deuterons up to 40 MeV with a current of 125 mA. The deuteron beam will strike a liquid lithium target, which circulates at high speed (15 m/s). It will reach a neutron flux of $10^{18}$ m$^{-2}$s$^{-1}$ with a broad peak at 14-20 MeV. Beyond the study of the materials to construct DEMO, more applications in several fields have been proposed at IFMIF-DONES [4].

In this paper, we perform a preliminary study on a new possible application: the production with deuterons of medical radioisotopes. Here, we study the production of $^{177}$Lu with the reactions named direct (i), and indirect (ii):
i) d+$^{176}$Yb →n+$^{177(m+g)}$Lu; ii) d+$^{176}$Yb →n+$^{177g}$Lu.

At present, $^{177}$Lu is used for theranostics (therapy and diagnosis) [5], or as a radiopharmaceutical to treat tumors as the gastroenteropancreatic neuroendocrine tumors [5]. Currently, $^{177}$Lu ($T_{1/2}$=6.65d) is only produced in nuclear reactors with neutron-induced reactions on Lu or Yb samples: $^{176}$Lu(n,ɤ)$^{177(m+g)}$Lu; and $^{176}$Yb(n,ɤ)$^{177}$Yb [6].

In this work, it has been considered a simple model for a production target of $^{177}$Lu. It consists of an Yb sample deposited onto a Cu backing cooled by flowing water. In the following, we will study the production of $^{177}$Lu with a $^{176}$Yb sample by means the two mentioned routes.

## MATERIALS AND METHOD

The boundary condition to calculate the $^{177}$Lu production will be to keep the temperature of the different parts of the target well below their melting points. The production of lutetium, through the direct route, is obtained by:

$$N_{177_{Lu}}(t) = \frac{I\left(1-e^{-\lambda_{177_{Lu}}t}\right)}{q\lambda_{177_{Lu}}} \int_{E_f}^{E_i} \frac{\sigma_{dir}(E)}{S_{Yb}^d(E)} dE; \quad (1)$$

For the indirect route, the production is obtained by:

$$N_{177_{Lu}}(t) = \frac{I}{q} \int_{E_f}^{E_i} \frac{\sigma_{ind}(E)}{S_{Yb}^d(E)} dE \left( \frac{1-e^{-\lambda_{177_{Lu}}t}}{\lambda_{177_{Lu}}} + \frac{e^{-\lambda_{177_{Yb}}t}-e^{-\lambda_{177_{Lu}}t}}{\lambda_{177_{Yb}}-\lambda_{177_{Lu}}} + \frac{1-e^{-\lambda_{177_{Yb}}t}}{\lambda_{177_{Lu}}-\lambda_{177_{Yb}}} \right); \quad (2)$$

Where $\sigma_{dir}(E)$ and $\sigma_{ind}(E)$ are the cross-section of the direct and indirect routes, and $S_{Yb}^d(E)$ is the stopping power calculated with SRIM [7]. For the values of the cross-section, the experimental data of Hermanne *et al* [8] and Manenti *et al* [9] have been used. The data range up only to 20 MeV, so for higher energies theoretical fits have been considered following [10], see Fig. 1.

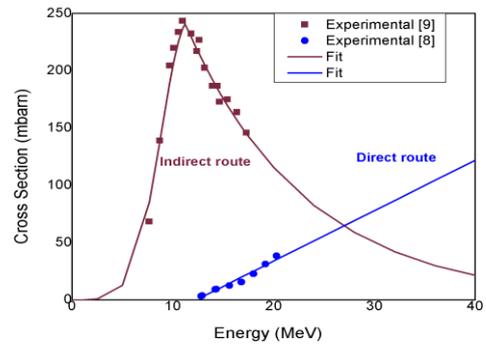

Fig. 1. Cross section (mbarn) of the reaction d+$^{176}$Yb, experimental data and theorical fit of both. The lines correspond to the fits following Arias de Saavedra *et al* [10].

Among others, the stopping power of deuterons and the thicknesses of the Yb and Cu determine the temperatures. The sample is attached to the backing, and this is in touch in the cooling fluid. Considering that the heat is transferred by conduction, along the target, and convection, due to the fluid in contact with the backing. In the approximation that the thickness of the target (l) is much lower than the radius (r), l << r, and stationary conditions, the heat dissipated by the fluid is:

$$q = h_t(T_{Yb} - T_f); \quad (3)$$

where $h_t$ is [11]:

$$h_t = \left( \frac{k_{Yb}2r\sqrt{\frac{0.5\rho v^2+101325}{\sigma_e}}+k_b l_{Yb}}{k_{Yb}k_b} + 27.03\frac{2r}{k_f}\left(\frac{\mu}{\rho v r}\right)^{0.8}\left(\frac{k_f}{c_p\mu}\right)^{0.43} \right)^{-1}; \quad (4)$$

where $k_f$ is the coefficient of thermal conductivity of the fluid, $\rho$ is the density of the water, $\mu$ is the dynamic viscosity of the fluid and $c_p$ is the specific heat at constant fluid pressure. We have considered a radius of 1 cm for the target and for the deuteron beam.

## RESULTS

In the following, we have considered a copper backing of 0.7 mm in thickness, and the water circulating at 5 m/s with a temperature of 20°C. The maximum deuteron current of 125 mA resulted in temperatures above the melting point for the feasible configurations. Thus, we have optimized the intensity to 1.25 mA to keep the temperature of the sample ($T_{Yb}$ = 1096 K) and the backing ($T_{Cu}$ = 1358 K) below their melting points, see table 1.

Table 1. Energy lost in Yb (ΔE) and temperature of the Yb foil and of the Cu backing for different ytterbium thicknesses ($l_{Yb}$) for 40 MeV and 1.25 mA deuteron beam. Similar results were obtained for $Yb_2O_3$ sample.

| ΔE (MeV) | $l_{yb}$ (mm) | $T_{Yb}$ (K) | $T_{Cu}$ (K) |
|---|---|---|---|
| 9.020 | 1 | 6492 | 5454 |
| 4.305 | 0.5 | 2998 | 2747 |
| 0.833 | 0.1 | 762 | 752 |
| 0.414 | 0.05 | 514 | 506 |
| 0.083 | 0.01 | 423 | 419 |

In case of nuclear reactors, the chemical form of the Yb sample is $Yb_2O_3$ with a maximum enrichment of 97% [12]. With the same enrichment, we calculate the production rate of $^{177}$Lu considering a foil thickness of 150 μm. Table 2 shows the results of the production rate of $^{177}$Lu.

Table 2. Values obtained for the production rate of $^{177}$Lu in milligram per second, $R_{Lu}$ (mg/s), with Yb sample and $Yb_2O_3$.

| $R_{Lu}$ (mg/s) | $^{196}$Yb | $^{196}Yb_2O_3$ |
|---|---|---|
| Direct route | 1.76·10$^{-7}$ | 3.08·10$^{-7}$ |
| Indirect route | 3.35·10$^{-8}$ | 6.03·10$^{-8}$ |

The higher production rate obtained for the $^{196}Yb_2O_3$ is due to it higher density than the natural ytterbium. With the same parameters, we have calculated the activity of lutetium after irradiation up to 24 hours of irradiation.

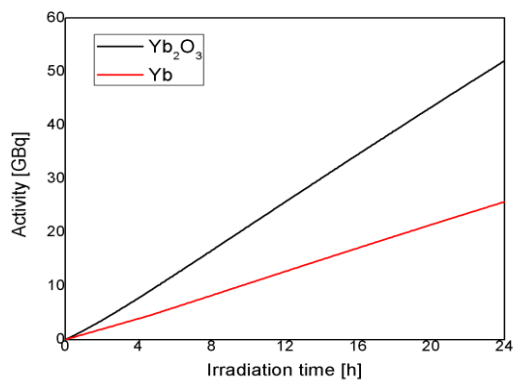

Fig. 2. Activity generated by $^{177}$Lu for $^{176}$Yb foil and $^{196}Yb_2O_3$ foil. Both, direct and indirect routes are included in the calculation.

## DISCUSSION AND CONCLUSION

We have studied the production of $^{177}$Lu used in therapy and imaging of cancers, through the two specific production routes, with 40 MeV deuterons at IFMIF-DONES. A simple target with cooling system has been studied by an analytical model. It has been shown that the deuteron intensity should be decreased from 125 mA to 1.25 mA. The specific activity obtained after 24 h at IFMIF-DONES is 0.119 GBq/mg for the $^{196}Yb_2O_3$ sample, and 0.076 GBq/mg for the $^{196}$Yb sample. We can compare these values to the specific activity produced in nuclear reactors with $^{196}Yb_2O_3$ sample after an irradiation of 72 h, which is 2.96 TBq/mg [12]. Although the specific activity is lower, the production at IFMIF-DONES would have a considerable impact in a regional health system as Granada (Spain). Conventionally a patient needs four doses during a treatment. Each dose has a price of 14 k€ per dose and an activity of 7.4 GBq per dose [13]. Therefore, in case of 24 hours of irradiation at IFMIF-DONES the produced $^{177}$Lu could save around 112 k€. It should be stressed that we are only providing preliminary calculations with the aim to motivate more realistic studies.

**Acknowlegments.** This work was carried out within the framework of the EUROfusion Consortium and has received funding from the Euratom research and training programme 2014-2018 and 2019-2020 under grant agreement No 633053. The views and opinions expressed herein do not necessarily reflect those of the European Commission. This work was supported by the Spanish projects FIS2015-69941-C2-1-P (MINECO-FEDER, EU), A-FQM-371-UGR18 (Programa Operativo FEDER Andalucia 2014-2020), the Spanish Association Against Cancer (AECC) (Grant No. PS16163811PORR), and the sponsors of the University of Granada Chair Neutrons for Medicine: Fundación ACS, Capitán Antonio and La Kuadrilla.